\documentclass[twocolumn,showpacs,aps]{revtex4} 
\usepackage{graphicx}
\usepackage{dcolumn}

\begin{document}

\title{An objective lens for efficient fluorescence detection of
single atoms}

\author{W. Alt}
 \email{w.alt@iap.uni-bonn.de}
 \affiliation{Institut f\"{u}r
 Angewandte Physik, Wegelerstrasse 8, 53115 Bonn, Germany}

\date{\today}

\begin{abstract}
We present the design of a diffraction limited, long working distance
monochromatic objective lens for efficient light collection.
Consisting of four spherical lenses, it has a numerical aperture of
0.29, an effective focal length of 36~mm and a working distance of
36.5~mm. This inexpensive system allows us to detect $8 \cdot 10^4$
fluorescence photons per second from a single cesium atom stored in a
magneto-optical trap.
\end{abstract}


\maketitle

\section{Introduction}

The optical detection of individual trapped particles, such as
single ions in a Paul trap~\cite{Neuh80} or neutral atoms in a
magneto-optical trap~\cite{Hu94}, necessitate the development of
efficient imaging optics. The ability to collect low levels of
fluorescence light and the imaging of small objects require a high
numerical aperture and diffraction limited performance,
respectively. Additionally a long working distance is often
necessary to provide access for laser beams or mechanical
structures to the trap region.

Commercially available solutions for this application are either long
working distance microscope objectives, which are relatively
expensive, molded aspheric lenses with usually short focal lengths,
or achromats, which have larger spot sizes. Special experimental
geometries, such as imaging through the window of a vacuum cell, or
specific geometrical restrictions, often require custom solutions.

\section{Lens design}

\begin{figure}
 \includegraphics[width=8.0cm]{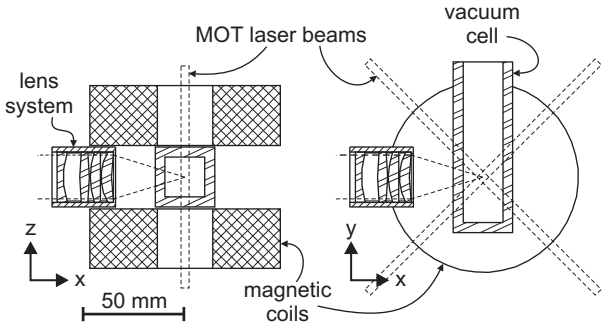}
 \caption{Experimental geometry}
 \label{setup}
\end{figure}

Our experimental requirement is the efficient collection of
fluorescence light from single atoms in a magneto-optical trap
(MOT)~\cite{Haub96}, see Fig.~\ref{setup}. The trap is located inside
a vacuum cell, which itself is situated between two magnetic coils.
This results in the following typical requirements for an objective
used for single atom experiments:

\begin{enumerate}

\item Collimation of the radiation of a point source at a
wavelength of 852~nm with a large numerical aperture (NA). Once the
light is collimated it can easily be imaged with low NA optics.

\item A minimum working distance of 36~mm. This is determined by
our laser setup because laser beams reflected off the glass cell must
not enter the objective. Moreover, the experimental geometry does not
permit any optical elements inside the vacuum cell.

\item Near diffraction limited spot size on axis, to enable
high resolution imaging and spatial filtering of the fluorescence
light.

\item A field of view covering the MOT position uncertainty of about
1~mm.

\item Imaging through a plane silica window of 5~mm thickness, which
introduces spherical aberrations for a NA$>$0.2.

\item Limitation of the outer diameter of the assembled lens system to 30~mm
by the magnetic coils of the MOT, see Fig.~\ref{setup}.

\end{enumerate}

To keep the design reasonably simple and cheap, we restricted
ourselves to spherical surfaces and BK7 glass only, and we used
standard catalog lenses whenever possible.

Single spherical lenses exhibit a reduction of their resolution above
a NA of about 0.1 due to their predominant spherical aberration. The
main idea of a multi-lens system is therefore to compensate for the
aberrations of one surface with the aberrations of other ones. In our
case, positive spherical aberration from convex surfaces is the
primary disturbance to eliminate. It needs to be balanced by negative
spherical aberrations of concave surfaces.

The objective is designed using the program Oslo LT~\cite{sinopt}. It
traces a parallel input beam of fixed diameter through the lens
elements and the 5~mm silica window. During the optimization the
radius of curvature of the last surface of the last lens is
controlled to keep the NA and effective focal length fixed. The
program is set up to minimize the squared sum of the spherical
aberrations up to 7th order and third order coma and astigmatism. The
radii of curvature and the lens distances are used as variables, and
the program´s optimization routine is iterated. Different starting
configurations are used in search for a global optimum.

\begin{figure}
 \includegraphics[width=8.0cm]{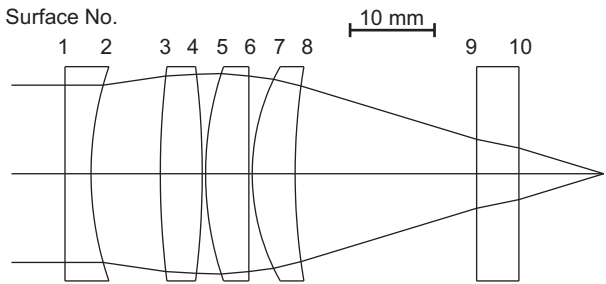}
 \caption{Layout of the lens system. Surfaces 1 to 8 represent the objective itself,
 surfaces 9 to 10 represent the silica wall of the vacuum cell.
 The MOT is at the position of the focus on the right.}
 \label{lenssyst}
\end{figure}

Acceptable performance could not be achieved with three lenses, with
four lenses, however, good designs were possible. The radii of
curvature of the lens surfaces were subsequently fixed to catalog
values of our vendor~\cite{lensoptics}, each time reoptimizing the
remaining variables. The resulting design is shown in
Table~\ref{datatable} and Fig.~\ref{lenssyst}. It consists of three
standard lenses and one meniscus lens with catalog radii of
curvature~\cite{Tamm}.

\begin{table}
 \caption{Specifications of the lens system.}
 \begin{ruledtabular}
 \begin{tabular}{cddc}
 \mbox{Surface} & \multicolumn{1}{c}{\mbox{Radius of}} & \multicolumn{1}{c}{\mbox{Distance to}} & \mbox{Material} \\
  \mbox{No.} & \multicolumn{1}{c}{\mbox{curvature [mm]}} & \multicolumn{1}{c}{\mbox{next surface [mm]}} & \\
  \hline
  1 & \multicolumn{1}{c}{$\infty$} & 3.08 & BK7 \\
  2 & 39.08 & 8.20 & air \\
  3 & 103.29 & 4.97 & BK7 \\
  4 & -103.29 & 0.40 & air \\
  5 & 39.08 & 5.12 & BK7 \\
  6 & \multicolumn{1}{c}{$\infty$} & 0.40 & air \\
  7 & 26.00 & 5.07 & BK7 \\
  8 & 78.16 & 21.55 & air \\
  9 & \multicolumn{1}{c}{$\infty$} & 5.00 & silica \\
  10 & \multicolumn{1}{c}{$\infty$} & 10.00 & vacuum \\
 \end{tabular}
 \end{ruledtabular}
 \label{datatable}
\end{table}

The design has a wavefront aberration of $\lambda/1000$ rms on axis,
resulting in a diffraction limited spot size of 1.8~$\mu$m (airy disc
radius). Provided that the curvature of the image surface is taken
into account, the wavefront error at a distance of 0.5~mm off axis is
$\lambda/13$ rms. 1~mm off axis the spot size radius increases to
3~$\mu$m rms.

Note that the performance of the objective is by no means limited to
the special requirements of our experimental setup. Changing the
distance between surfaces 2 and 3 in Fig.~\ref{lenssyst} from 8.2~mm
to 6.6~mm (and refocusing) allows the system to work even without the
5~mm silica window, with negligible performance degradation. Adaption
to any window thickness up to 10~mm is possible.

Although the design is optimized for 852~nm it retains its
diffraction limited performance from 1064~nm to 400~nm when the
chromatic focus shift is taken into account.

\section{Assembly and experimental tests}

All lenses have a diameter of 1", a surface quality (over 90~\% of
the clear aperture) of $\lambda/4$, scratch-dig 20-10, centration $<$
5 minutes of arc and AR-coating for 650-1000~nm. They are stacked
into an aluminum tube of 1" inner and 30~mm outer diameter and held
in place by a threaded retainer ring. The distances between the
lenses are determined by thin aluminum spacer rings. The mechanical
parts have been manufactured to a tolerance of 0.1~mm.

\begin{figure}
 \includegraphics[width=8.0cm]{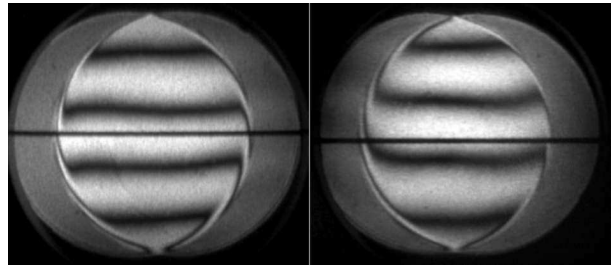}
 \caption{Shear plate interferograms of the beam produced by
 collimating a point source with the objective lens.
 The shear is applied in two orthogonal directions,
 straight equidistant interference fringes correspond to a plane wavefront.}
 \label{shearimage}
\end{figure}

An experimental test of the wave front aberration was performed by
focusing an 852~nm laser beam onto a 1 $\mu$m diameter pinhole
serving as a high NA point source. The transmitted light was
collimated by the objective under test and analyzed by means of a
shear plate interferometer~\cite{shearplate}. From the bending of the
resulting interference fringes (Fig.~\ref{shearimage}) we estimate a
wavefront distortion of less than $\lambda/4$ peak-valley over 90~\%
of the clear aperture.

In our MOT setup (Fig.~\ref{setup}) the optical axis of the objective
was carefully aligned onto the trap center using a Helium-Neon laser
beam. The collimated fluorescence light from the MOT is focused by an
f=80~mm doublet lens through a pinhole of 150~$\mu$m diameter for
spatial filtering (stray light suppression). The transmitted light is
imaged onto an avalanche photodiode operated in single photon
counting mode, with a quantum efficiency $\eta \approx 50$~\% at
852~nm. When the MOT lasers are saturating the atom, we detect about
$8 \cdot 10^4$ fluorescence photons per second from one cesium atom
on a stray light background of only $2.2 \cdot 10^4$ photons per
second. Since with a NA of 0.29 we cover $\Omega/(4\pi) = 2.1$~\% of
the total solid angle, and the lifetime of the excited state is $\tau
= $30.5~ns, the theoretical upper limit for the count rate of a
strongly saturated two level atom is

\begin{equation}
 R = \eta \frac{\Omega}{4\pi} \frac{1}{2\tau} = 17 \cdot
 10^4 \mbox{s}^{-1} .
\end{equation}

Due to the more complex situation of a cesium atom in the
MOT~\cite{Townsend95} the discussion of which is beyond the scope of
this paper the expected value is significantly below the upper limit.

\section{Conclusions}

Using standard lenses we have successfully designed and built a
diffraction limited, long working distance lens system for collecting
fluorescence from single atoms in a MOT. The total cost of the
objective including lenses and mechanical parts is about US~\$~500
only. A similarly inexpensive system could be advantageously used for
high resolution imaging of single ions or atoms or for heterodyne
detection of fluorescence radiation. This system recently managed to
detect single atoms in an optical dipole trap for the first
time~\cite{Kuhr2001}.

\begin{acknowledgments}
 We thank M. Dornseifer, Dr. C. Tamm, Dr. S. Kanorsky and M. Schulz
 for discussions. This work was supported by the Deutsche
 Forschungsgemeinschaft.
\end{acknowledgments}


\begin{thebibliography}{99}
\bibitem{Tamm}
Dr. C. Tamm has independently designed a similar lens system for the
observation of single ions through a vacuum window (private
communication).
\end{thebibliography}
\end{document}